\documentclass[twocolumn,showpacs,preprintnumbers,ams
m a t h , a m s symb]{revtex4} 
\usepackage[dvips]{graphicx}
\usepackage{verbatim}
\begin{document}

\preprint{x}

\author{A C Graham}
\affiliation{Cavendish Laboratory, Madingley Road,
Cambridge, CB3 OHE, United Kingdom}

\author{M Pepper}
\affiliation{Cavendish Laboratory, Madingley Road,
Cambridge, CB3 OHE, United Kingdom}

\author{M Y Simmons$^{\ast}$}
\affiliation{Cavendish Laboratory, Madingley Road,
Cambridge, CB3 OHE, United Kingdom}

\author{D A Ritchie}
\affiliation{Cavendish Laboratory, Madingley Road,
Cambridge, CB3 OHE, United Kingdom}

\title{Anomalous spin-dependent behaviour of one-dimensional subbands}

\date{\today}

\begin{abstract}
We report a new electron interaction effect in GaAs/AlGaAs quantum wires. Using DC-bias spectroscopy, we show that large and abrupt changes occur to the energies of spin-down (lower energy) states as they populate.  The effect is not observed for spin-up energy states.  At B=0, interactions have a pronounced effect, in the form of the well-known 0.7 Structure.  However, our new results show that interactions strongly affect the energy spectrum at \textit{all} magnetic fields, from $0$ to $16~$T, not just in the vicinity of the 0.7 Structure.

\end{abstract}

\maketitle

Semiconductor nanostructures such as quantum wires are involved in the development of future quantum technologies, so it is important that their electronic properties are understood.  Ballistic quantum wires are ideal quantum laboratories for studying many-body physics, due to the low electron densities combined with the low levels of disorder which can now be achieved. However, despite their simple geometry, the properties of these quasi-one-dimensional (Q1D) systems are much less well understood than those of 2D or 0D systems. An example of this is the 0.7 Structure \cite{thomas96}, a spin-related phenomenon which has attracted much interest in recent years, although its exact origin is still the subject of debate.  It has recently been discovered that the 0.7 Structure is not an isolated phenomenon, but has a high-magnetic field variant which occurs at crossings of Zeeman-split 1D subbands\cite{abiprl}.

  In this paper, we report a new interaction effect which demonstrates that electron-electron interactions profoundly affect the electronic structure of the quantum wire at \textit{all} magnetic fields, not just in the region of the 0.7 Structure or its high-field version, the Analog \cite{abiprl}.  Using DC-bias spectroscopy \cite{patel91b}, we show that as spin-down (lower energy) subbands begin to populate, they abruptly drop in energy by as much as $0.5~$meV, within a vanishingly small range of gate-voltage. These results also cast doubt on basic assumptions which are often made when analysing quantum-wires - such as the idea that electron density is proportional or equivalent to gate-voltage, and the validity of an `exchange-enhanced g-factor' in quantum wires \cite{thomas98}.

Our new findings also provide an explanation for one of the most characteristic features of the 0.7 Structure, which has, to our knowledge, not been explained by any model or theory so far - the question of how the 0.7 Structure can survive to such high temperatures, when all other conductance features have disappeared \cite{thomas98}.

Our samples consist of split-gate devices \cite{Thornton} defined by electron-beam lithography on a Hall bar etched from a GaAs/Al$_x$Ga$_{1-x}$As heterostructure. All the samples used in this work have a length of 0.4~$\mu$m and a width of 0.7~$\mu$m. The two-dimensional electron gas (2DEG) formed $292\; $nm below the surface has a mobility of $1.1\times 10^{6} \;$cm$^{2}$/Vs and a carrier density of
$1.15\times 10^{11}\;$cm$^{-2}$.  In the parallel magnetic field regime, using the Hall voltage, the out-of-plane misalignment was measured to be $0.3\,^{\circ}$. The measurement temperature was $50~$mK.

\begin{figure}
\begin{center}
\includegraphics[width=0.85\columnwidth]{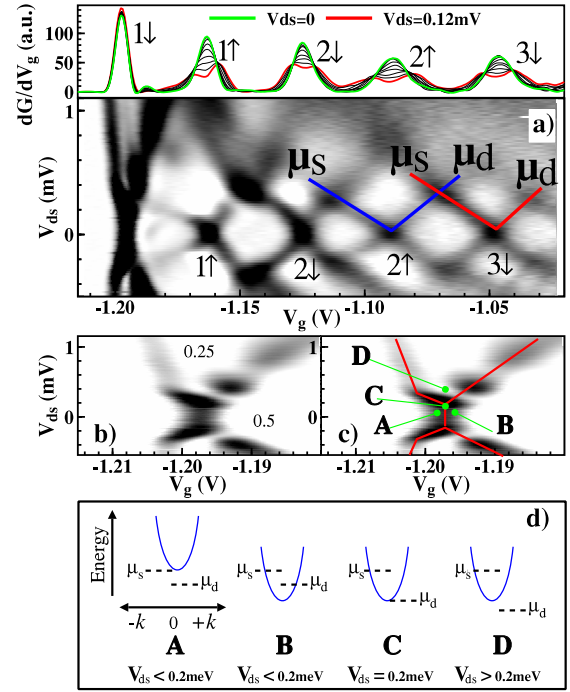}
\end{center}
\caption{(a) Grey-scale of $dG/dV_{\rm{g}}$ data at $5~$T as a function of $V_{\rm{ds}}$ and $V_{\rm{g}}$. Upper panel:  $dG/dV_{\rm{g}}$ at $5~$T from $V_{\rm{ds}}=0$ (top trace) to $V_{\rm{ds}}=0.12~$mV (bottom trace). (b) Close-up of features in (a) between $V_{\rm{g}}=-1.18~$V and $V_{\rm{g}}=-1.215~$V. Conductances of the plateau regions are marked in units of $2e^2/h$. (c) As (b) but annotated. (d) Configuration of energy levels at the points marked in (c). \label{Fig2}}
\end{figure}

\begin{figure}
\begin{center}
\includegraphics[width=0.85\columnwidth]{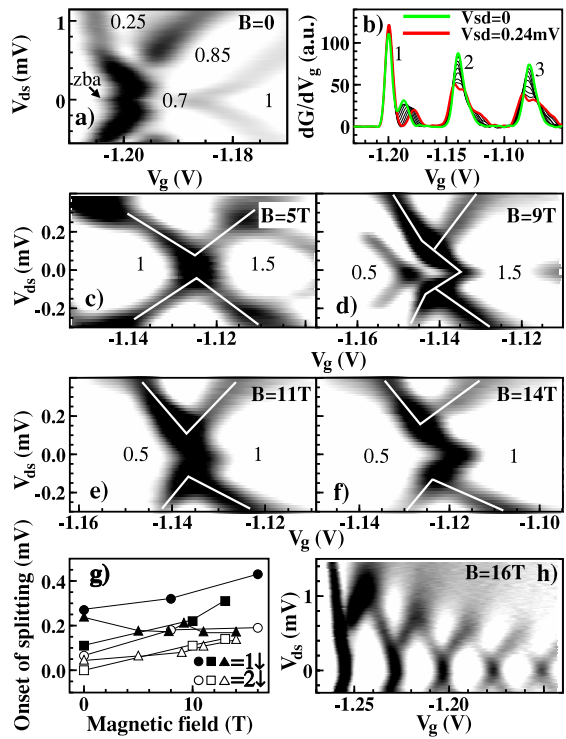}
\end{center}
\caption{(a) A typical example of $V_{\rm{ds}}$ data for the first subband at $B=0$. (b) $dG/dV_{\rm{g}}$ data used in (a), from $V_{\rm{ds}}=0$ to $0.24~$mV in steps of $0.03~$mV, for the first three subbands.  The pinch-off voltages have been aligned for clarity, and the peaks labelled with their corresponding subbands. (c)-(f) $V_{\rm{ds}}$ data for 2$\downarrow$ at $B=5,9,11 \& 14$T.  V-shaped splitting begins at $0.07~$mV, $0.09~$mV, $0.12~$mV and $0.15~$mV respectively. The white lines are a guide to the eye. (g) Plot of the biases at which splitting begins, as a function of $B$, for 1$\downarrow$ and 2$\downarrow$ (open or closed symbols) of three samples (circles, squares or triangles). The onset of splitting is found by marking the $dG/dV_{\rm{g}}$ peak-trajectories by hand and finding their intersection.(h) $V_{\rm{ds}}$ data from another sample at $16~$T. V-shaped splitting on the first two subbands clearly begins at $V_{\rm{ds}}>0$. \label{Fig3}}
\end{figure}

Most of the data in this paper is in the form of grey-scale diagrams, where the transconductance, $dG/dV_{\rm{g}}$ (where $G$ is the differential conductance), is plotted as a function of split-gate voltage $V_{\rm{g}}$, and DC-bias, $V_{\rm{ds}}$ \footnote{$dG/dV_{\rm{g}}$ is the numerical differential of $G$.  In all our figures, the step-size in $V_{\rm{ds}}$ is $0.015~$mV, and in $V_{\rm{g}}$ is $800~\mu~V$.}.  To interpret the features of the data, we must consider what would be expected for finite $V_{\rm{ds}}$ for a system of non-interacting spin-split 1D subbands \cite{glazman}.  This model can be generalised in a trivial fashion for spin-degenerate subbands.  At $V_{\rm{ds}}=0$, the conductance will rise by $e^2/h$ when a subband passes through the chemical potential.  For finite $V_{\rm{ds}}$, the electrochemical potential is split in two, where $\mu_s=\mu+eV_{\rm{ds}}/2$ and $\mu_d=\mu-eV_{\rm{ds}}/2$ are the electrochemical potentials of source and drain respectively.  $\mu_{\rm{s}}$ and $\mu_{\rm{d}}$ are separated by a known energy, $eV_{\rm{ds}}$.  In this case, the conductance will increase by only $e^2/2h$ when the subband passes through $\mu_{\rm{s}}$, and by another $e^2/2h$ when the subband subsequently passes through $\mu_{\rm{d}}$.  It will require a finite change in gate voltage to shift the subband from $\mu_{\rm{s}}$ to $\mu_{\rm{d}}$.  This gate-voltage difference will increase for larger $V_{\rm{ds}}$ as $\mu_{\rm{s}}$ and $\mu_{\rm{d}}$ move apart in energy.

When plotted in grey-scale form, the peaks in $dG/dV_{\rm{g}}$ at zero bias will split into two at finite bias, with the left-hand peak of the pair corresponding to the interception of the subband with $\mu_{\rm{s}}$ and the right-hand peak corresponding to interception with $\mu_{\rm{d}}$.  The trajectories of these peaks as a function of gate voltage, $V_{\rm{g}}$, and of $V_{\rm{ds}}$ will form a V-shape.  Clearly, both branches of the V-shape should be present, as the subband must pass through both electrochemical potentials.  Furthermore, the onset of the V-shaped splitting should occur as soon as $V_{\rm{ds}}>0$.

We will now apply these concepts to our data, but rather than starting with $B=0$, we will first consider $B=5~$T.  The presence of the 0.7 Structure at $B=0$ complicates the analysis, as the exact configuration of the energy spectrum is not well established.  However, at a field of $5~$T, the 0.7 Structure has evolved into a plateau at $e^2/h$ \cite{thomas96,abiprl}, and the zero-bias conductance characteristics are well-described by a non-interacting model, making $5~$T the simplest case to consider.

The $V_{\rm{ds}}$ characteristics at $5~$T are shown in the lower panel of fig.~\ref{Fig2}(a).  Regions of large $dG/dV_{\rm{g}}$, corresponding to risers, are plotted in black and regions of low $dG/dV_{\rm{g}}$, corresponding to plateaux, are plotted in white.  From the left, the second white region centred on $V_{\rm{ds}}=0$ corresponds to the $e^2/h$ plateau, and the third region, lying between $V_{\rm{g}}=-1.12$ and $-1.17~$V, corresponds to the plateau at $2e^2/h$.  The black V-shaped features associated with subbands 2$\uparrow$ and 3$\downarrow$ are marked, and the branches are labelled indicating whether they correspond to the subband intercepting $\mu_{\rm{s}}$ or $\mu_{\rm{d}}$.   
    
The left-most V-shape centered on $V_{\rm{g}}=-1.195~$V is associated with the spin-down level from the first subband, 1$\downarrow$.  Looking at the close-up of this feature in \ref{Fig2}(b), it is clear that it is not the simple V-shape we expected.  The V-shaped splitting does not begin until around $0.2~$meV, before which there is only one dark branch.  This absence of splitting is \textit{not} simply a case of two closely spaced peaks being poorly resolved due to their finite width --- even in this case, the peak should broaden as soon as a bias is applied:  the upper panel of fig.~\ref{Fig2}(a) shows $dG/dV_{\rm{g}}$ traces used for the grey-scale, from $V_{\rm{ds}}=0$ (the upper trace) to $0.12~$meV (the lowest trace) in $0.015~$meV steps.  Even at $0.12~$meV, splitting is resolvable on the 1$\uparrow$ and 2$\downarrow$ peaks, and in addition, apart from the 1$\downarrow$ peak, all others broaden as soon as the bias is applied, i.e. at the lowest increment of $V_{\rm{ds}}=0.015~$meV.  From zero to $0.12~$meV, the full-width at half-maximum of the 1$\uparrow$ peak has increased by 115\%, and for 2$\downarrow$ by 117\%.  In contrast, by $0.12~$meV, the  1$\downarrow$ peak has broadened by less than 7\%, its transconductance has \textit{increased}, and it exhibits no splitting \footnote{The very small peak at the base of $1\downarrow$ is a small resonance on the $e^2/h$ plateau caused by disorder --- it appears in the greyscale as a small smudge around $V_{\rm{ds}}=0$.  It is not a universal feature, so not significant.}. 

This behaviour is a clear deviation from the simple non-interacting model. The finite-bias conductance changes continuously from zero to $e^2/h$ with no features in between --- it appears that the 1$\downarrow$ subband has managed to pass through both $\mu_{\rm{s}}$ and $\mu_{\rm{d}}$ within a very narrow gate voltage range, despite the fact that $\mu_{\rm{s}}$ and $\mu_{\rm{d}}$ are separated in energy by $eV_{\rm{ds}}$.  We propose the following interpretation:  this behaviour indicates that 1$\downarrow$ rapidly drops in energy as soon as it begins to populate, abruptly dropping through both electrochemical potentials.  Since the V-shaped splitting begins at $V_{\rm{ds}}=0.2~$mV, this implies that 1$\downarrow$ instantaneously drops in energy by $0.2~$meV when it intercepts $\mu_{\rm{s}}$.  Figure \ref{Fig2} (c) and (d) illustrate this.  At point \textbf{A}, 1$\downarrow$ has just intercepted $\mu_{\rm{s}}$.  Within a very small change in  $V_{\rm{g}}$, 1$\downarrow$ has dropped a long way in energy as shown in \textbf{B}.  Because it has dropped by more than the separation of $\mu_{\rm{s}}$ and $\mu_{\rm{d}}$,  the conductance increases directly from zero to $e^2/h$ (corresponding to point \textbf{B}), and no splitting in $dG/dV_{\rm{g}}$ occurs --- we cannot resolve this subband populating at $\mu_{\rm{d}}$, as it has dropped through $\mu_{\rm{d}}$ almost instantaneously after it has dropped through $\mu_{\rm{s}}$.  If $V_{\rm{ds}}$ is increased to around $0.2~$meV, then two $dG/dV_{\rm{g}}$ peaks begin to appear.  Thus, the separation of $\mu_{\rm{s}}$ and $\mu_{\rm{d}}$ must now be slightly larger than the drop in energy experienced by the 1$\downarrow$ subband, corresponding to point \textbf{C}.  Although the two peaks in $dG/dV_{\rm{g}}$ are resolvable for $V_{\rm{ds}}>0.2~$meV, they are still very closely spaced in $V_{\rm{g}}$.  This suggests that at this bias, 1$\downarrow$ still drops rapidly in energy after intercepting $\mu_{\rm{s}}$, but this time, it jumps to a position \textit{just above} $\mu_{\rm{d}}$.  Eventually, at point \textbf{D}, there is a well-pronounced plateau region with a conductance of $0.25(2e^2/h)$, which corresponds to only negative $k$-states in 1$\downarrow$ being populated, as it now lies between $\mu_{\rm{s}}$ and $\mu_{\rm{d}}$.

The same behaviour also occurs for the first subband at $B=0$.  Figure \ref{Fig3}(a) shows a close-up of $dG/dV_{\rm{g}}$ data as a function of $V_{\rm{g}}$ and $V_{\rm{ds}}$ for $B=0$. The conductances of plateau regions are marked for clarity, in units of $2e^2/h$. Comparing this to fig.~\ref{Fig2}(b) at $5~$T, we note two main differences: firstly, there is a so-called `Zero-bias anomaly', labelled \textbf{zba}, in the $B=0$ data, not present in the $5~$T data. This is thought to indicate that Kondo physics can occur in quantum wires at $B=0$ \cite{marcus}, although it is probably not the cause of the 0.7 Structure  \footnote{Here, we briefly remark that we have not observed zero bias anomalies in the conductance anywhere in the region of crossings at high $B$, although we do observe structures enhanced in conductance by almost $e^2/h$, known as 0.7 Analogs. We deduce that Kondo-like physics is not the cause of this conductance enhancement and resultant Analogs.  Therefore we believe that although Kondo-physics may well occur at $B=0$, it is not the root cause of the 0.7 Structure, although it may enhance its conductance.}.  A second difference is that the pale grey branch in fig.~\ref{Fig3}(a) separating the $0.7$ and $0.85$ regions from the $1(2e^2/h)$ region, shifts to more positive $V_{\rm{g}}$ with increasing $B$ --- in fig.~\ref{Fig2}(a) at $5~$T it has evolved into the branch dividing the $e^2/h$ plateau from $2e^2/h$, corresponding to $1\uparrow$.  Importantly however, there is a major similarity between figs~\ref{Fig3}(a) and \ref{Fig2}(b). In fig.~\ref{Fig3}(a), the left-most dark feature exhibits no splitting in DC bias until $V_{\rm{ds}}>0.2~$mV, just as for $5~$T.  To make this clear, we show the $dG/dV_{\rm{g}}$ traces in fig.~\ref{Fig3}(b) for the three lowest subbands.   By $V_{\rm{ds}}=0.24~$mV, the left-most peak has broadened by only $3\%$, its transconductance has \textit{increased}, and no splitting is apparent, unlike for the other subbands.  This data is very similar to the upper panel of fig.~\ref{Fig2}(a). Furthermore, the V-shaped feature in fig.~\ref{Fig2}(b) at $5~$T evolves continuously from the V-shaped feature at $B=0$ in fig.~\ref{Fig3}(a), the only qualitative change being the disappearance of the zero-bias anomaly.  This is compelling evidence that the delay in the onset of V-shaped splitting has the same origin for both zero and finite magnetic fields.

We now present data showing that the absence of V-shaped splitting in a DC-bias is not unique to the first subband.  Close-ups of the $V_{\rm{ds}}$ data for 2$\downarrow$ are shown for $B=5$ through to $14~$T in figs~\ref{Fig3}(c) to (f).  Again, the plateaux are marked with their conductances in units of $2e^2/h$.  Interestingly, as $B$ increases, the onset of the V-splitting moves progressively to higher $V_{\rm{ds}}$ --- at $5~$T the splitting starts at $0.07~$meV, and by $14~$T this has increased to $0.15~$meV.   The bias at which splitting begins is plotted in fig.\ref{Fig3}(g) for 1$\downarrow$ and 2$\downarrow$ (open or closed symbol) for three samples (represented by three symbols).  This bias increases \footnote{The data labelled with filled triangles does not follow the trend of the other samples, but is plotted for completeness.  This discrepancy is because this sample had a resonant structure on the lowest plateau, caused by disorder.} with $B$, which implies that the subbands drop further in energy at higher fields. Fig. \ref{Fig3}(h) shows $V_{\rm{ds}}$ data taken at $16~$T from a sample in which the 1$\uparrow$ subband has crossed five times by this field.  Thus, \textit{all} of the features in fig.~\ref{Fig3}(h) are associated with spin-down subbands --- from the left of the figure, these are 1$\downarrow$, 2$\downarrow$, 3$\downarrow$, 4$\downarrow$ and 5$\downarrow$ respectively.  For 1$\downarrow$, 2$\downarrow$ and 3$\downarrow$, the splitting in $V_{\rm{ds}}$ is strongly offset from zero, with the effect becoming less pronounced with increasing density and subband index.

Such behaviour is not observed for spin-up subbands at any magnetic field: any offset to the V-shaped splitting for spin-up subbands is either too small to measure in our data, or is absent.  This indicates that spin-up subbands drop more gradually in energy, so that the energy gap between $\mu_{\rm{s}}$ and $\mu_{\rm{d}}$ is resolvable, in the form of splitting.  This may explain our earlier observation \cite{abipe} that for $B>0$, $dG/dV_{\rm{g}}$ peaks are larger for spin-down than for spin-up subbands.  This is consistent with the spin-down subbands moving rapidly through the chemical potential --- the associated change in conductance would be more rapid as a function of gate voltage, resulting in steeper risers between plateaux, and hence, larger $dG/dV_{\rm{g}}$.

The rapid population of spin-down subbands also has important implications regarding the 0.7 Structure.  Previous work \cite{kristensenbruus,reillyprl,abiprl,abissc,karl05} provides strong evidence that the 0.7 Structure is related to the opening of an energy gap between spin-up and spin-down energy levels as they populate.  Our observation that spin-down, but not spin-up levels, drop rapidly in energy as they populate, implies that the spin gap must open almost instantaneously to as much as $0.5~$meV, as the first subband is populated.  This is not in agreement with the model in reference \cite{reillyprl} where it is assumed that a spin gap opens gradually and linearly with increasing gate-voltage.  Our results also give insight into the temperature ($T$) dependence of the riser up to the 0.7 Structure.  It has been observed that the 0.7 Structure is still present at $T$ high enough to smear out all other conductance features\cite{thomas96}.  In fact, it is the \textit{riser} up to the 0.7 Structure that is almost invariant with temperature, which gives the impression that the 0.7 Structure does not experience significant thermal smearing --- at high $T$, above this riser, there are no other features in the conductance (see for example ref.~\cite{thomas98}, fig.~3(a)). This surprising behaviour is entirely consistent with 1$\downarrow$ dropping abruptly in energy as it populates. The bottom of the subband would rapidly jump through a thermally broadened chemical potential within a small $V_{\rm{g}}$ range, to a region of the Fermi distribution with nearly full occupancy.  The effects of the broadened chemical potential would therefore not be observed until much higher $T$ than would be expected in a non-interacting picture.

We note that if the energy of the spin-down subband changes nearly discontinuously as it populates, then the electron density and electrostatic potential in the quantum wire must also change abruptly as a function of $V_g$.  Therefore, in the strongly interacting regime at low densities, for example in the region of the 0.7 Structure,  $V_g$ and electron density are not equivalent.

Our results are also not consistent with the conclusions of reference \cite{picciotto} in which it is suggested that in, for example, fig.~\ref{Fig3}(a), the grey boundary between the $0.85(2e^2/h)$ region and $1(2e^2/h)$ corresponds to the first subband passing through $\mu_{\rm{d}}$, therefore corresponding to a transition from uni-directional to bi-directional transport in the wire.  As our data in a magnetic field demonstrates (see fig.~\ref{Fig2}(b)), the right-moving dark line on the right-hand side of the $0.25$ region corresponds to the 1$\downarrow$ subband passing through the drain electrochemical potential $\mu_{\rm{d}}$.  This dark line is also present in the data at $B=0$, therefore in the $0.85$ region, the transport is already bi-directional for 1$\downarrow$.

In the remainder of this paper we shall consider whether any existing theories can explain our results.  As we have stated,  we only observe an abrupt drop in energy for spin-down subbands.  This effect is therefore probably an exchange-interaction effect.  Wang and Berggren have demonstrated that at $B=0$, the 1D subbands in an infinite quantum wire spontaneously spin-split as they pass through $\mu$, due to exchange interactions \cite{wang96}.  The spin gap opens abruptly with a small change in the electron density, with the spin-down energy level dropping markedly below $\mu$. It was later shown that this also occurs for short quantum wires, although in this case the spin-gap opens slightly less abruptly \cite{karl02}. 

This essential mechanism qualitatively agrees with our results.  Furthermore, in both the theory and our experiments, the opening of the spin gaps and the related drop in energy of the spin-down levels occur in higher subbands as well as the first, with the effect diminishing with increasing subband index and electron density.  However, an exact understanding of our data is likely to be theoretically challenging.  We have observed that the V-shaped splitting induced by a source-drain voltage moves to progressively higher biases with increasing $B$.  This implies that the exchange-enhanced spin-splitting increases with $B$, which is consistent with the fact that crossings of spin-split subbands \cite{abiprl} occur at significantly lower magnetic fields than would be expected, given the bulk GaAs $g$-factor of -0.44.  In addition to this effect of Zeeman splitting on the exchange interaction, magnetic confinement effects may also play a role. At present it is unclear whether the $B$ dependence of the V-shaped splitting can be understood in the context of existing theories.

To conclude, we have studied the transport properties of short one-dimensional constrictions as a function of DC bias, over a range of magnetic fields. The data exhibits a new spin-related many-body interaction effect. It is observed that for spin-down subbands, the V-shaped splitting of peaks in $dG/dV_{\rm{g}}$ is absent until DC biases of as much as $0.5~$mV.  Such an effect is not observed for spin-up subbands.  We have demonstrated that this is not due to thermal broadening, and interpret the behaviour as strong evidence that spin-down subbands drop abruptly in energy as they begin to populate, as a result of electron interactions in the Q1D constriction. The large and abrupt changes in the energy spectrum as spin-down subbands populate also indicate that electron density in the Q1D channel changes abruptly as a function of gate-voltage.  In addition, this phenomenon provides an explanation for the high-temperature behaviour of the 0.7 Structure \cite{thomas98}, and other previously observed spin-asymmetries \cite{abipe}.

We thank K.-F. Berggren, P. Jaksch, M. Colarieti-Tosti, C. J. B. Ford, C. H. W. Barnes, K. J. Thomas, D. L. Sawkey and V. Tripathi for useful discussions. We also acknowledge the COLLECT European Research Training Network. This work was supported by EPSRC, UK. ACG acknowledges support from Emmanuel College, Cambridge.

$^{\ast}$Current address: University of New South Wales, School of Physics, Sydney, NSW 2052, Australia.
\bibliography{refs}

\end{document}